\documentstyle[multicol,epsf,bm,prl,aps]{revtex}

\begin{document}
\newcommand{\tr}{\mathop{\mathrm{tr}}\nolimits}
\newcommand{\wideeqtop}{%
\vskip-3.5mm
\noindent
  \vrule width3.4in height.2pt depth.2pt %
  \vrule depth0em height2mm \hfill
\vspace*{1.5mm}
}
\newcommand{\wideeqend}{%
\vskip1mm
\indent
  \hfill\vrule depth2mm height0pt %
  \vrule width3.4in height.2pt depth.2pt
\vskip0mm
}
\newcommand{\referencetop}{%
\begin{center}
\vrule width83mm height.2pt depth.2pt\\
\medskip
\end{center}
}

\draft
\preprint{}
\title{Flow equation approach to diagonal representation of \\
an unbounded Hamiltonian with complex eigenvalues}
\author{%
Yukiko Ohira and
Kentaro Imafuku$^1$\thanks{Email: imafuku@volterra.mat.uniroma2.it}%
}

\address{%
Department of Physics, Waseda University, Tokyo 169-8555,
Japan\\
Centro Vito Volterra, Universit\`a degli Studi di
Roma Tor Vergata, Roma 00133, Italy$^1$
}
\date{
\today
}
\maketitle
\begin{abstract}
The flow equation approach investigated by Wegner {\it et al}. 
\cite{ref:Wegner,ref:Lenz-Wegner,ref:Kehrein-Mielke,ref:Mielke}
is applied to an unbounded Hamiltonian system with a generalization.  
We show that a well-known quantized complex energy eigenvalues 
which is related to decay widths can be given with  this approach.
\end{abstract}

\begin{multicols}{2}
\narrowtext
To investigate a quantum system, a diagonalization of a given 
Hamiltonian is one of mighty approaches.  Eigenvalues 
and eigenstates which can be obtained with the diagonalized 
representation  are most important clues 
to understanding of the physical properties of the system.  
But unfortunately, however there are a few explicitly solved 
exceptions, an analytically diagonalization of a given Hamiltonian is 
hard problem and various schemes have been deviced. Among of them, 
a novel approach to this problem was proposed by Wegner \cite{ref:Wegner}. 
A way to construct of a generator, which generates one parameter unitary 
group which drives a given initial Hamiltonian to diagonalized 
(or almost diagonalized) final one, was deviced.  
This approach is called flow equation approach and it 
has already been applied to various physical models, especially  
condensed matter physics and high energy physics with some techniqal 
improvements\cite{ref:Lenz-Wegner,ref:Kehrein-Mielke,ref:Mielke}.  

On the other hand,  there is interesting argument on 
the diagonalization, which is the diagonal representation  
of a Hamiltonian unbounded from below with complex eigenvalues.
The imaginary part of the eigenvalues is related to the life-time of 
the quasi-stationary state, and it plays a good role to investigate 
a resonance phenomena\cite{ref:Gamov,ref:Nakanishi,ref:Combes,ref:Bohm,ref:Shinbori-Kobayashi,ref:Shinbori}, however the eigenfunctions belonging 
to the complex eigenvalues can not be interpreted in a conventional sense. 
A mathematical treatment of the quasi-stationary approach 
has been also studied with a rigged Hilbert space 
(or Gel'fand triplet) formulation of quantum mechanics
\cite{ref:Bohm,ref:Shinbori-Kobayashi,ref:Shinbori}. 
As pointed out by Bohm {\it etal} \cite{ref:Bohm}, such formulation 
allows for the description of the {\it intrinsic irreversible processes} in 
quantum mechanics, and these studies would be important in order to understand 
the fundamental relation between macroscopic irreversible phenomena 
and microscopic reversible theory.

Our interest is an investigation or generalization of Wegner's 
flow equation to the unbounded system problems. 
In this letter, we consider one-dimensional quadratic interaction model 
(\ref{eq:hamiltonian}), 
which becomes unbounded  below in cases of $|\lambda_0/\omega_0|>1/2$. 
We show that the flow equation approach can be generalized to such situation 
where the original approach does not work, and this treatment provides 
a good scheme to study such deep features as a time arrow in quantum mechanics
which were discussed in \cite{ref:Bohm,ref:Shinbori-Kobayashi,ref:Shinbori}.

We shall consider a quadratic interaction system with the following Hamiltonian
\begin{equation}\label{eq:hamiltonian}
H=\omega_{0} ~a^{\dagger}a+\lambda_{0} 
~\left({a^{\dagger}}^2+a^2\right)+v_{0}
\end{equation}
where $\omega_{0}$, $\lambda_{0}$ and $v_{0}$ are given constant of 
real numbers.

\begin{figure}
\epsfxsize=0.4\textwidth
\epsfbox{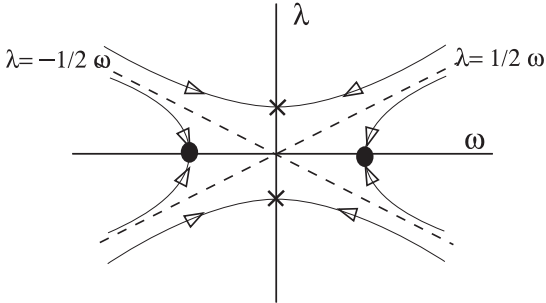}
\caption{Schematical illustration of the unitary flow. 
The filled circles and crosses represent the fixed points
(\ref{eq:bounded_case}) and (\ref{eq:fixed_point_2}).
\label{fig:unitary_flow}}
\end{figure}

In order to explain the wegner's approach and our interest, 
first, let us apply the  original flow equation approach 
to this Hamiltonian (\ref{eq:hamiltonian}). 
The flow equation is a continuous unitary transformation
of the Hamiltonian, which is written in the following differential form;
\begin{mathletters}
\label{eq:wegner_equation}
\begin{equation}
\frac{dH(l)}{dl}=[\eta(l),H(l)], \quad
\eta(l)=[H_d(l),H(l)]
\end{equation}
or
\begin{equation}\label{eq:U_transformation}
H(l)=U^{\dagger}(l)H(0)U(l),
\quad\frac{d}{dl}U^{\dagger}(l)=\eta(l)U^{\dagger}(l)
\end{equation}
\end{mathletters}
with the initial condition 
$H(0)=H$ (original (given) Hamiltonian). 
$H_{d}(l)$ is  the diagonal portion of $H(l)$.
Applying to the Hamiltonian (\ref{eq:hamiltonian}), 
we obtain the following set of 
differential equations by coefficient matching:
\begin{mathletters}\label{eq:flow_equation}
\begin{eqnarray}
\frac{d\omega(l)}{dl}&=&-16\omega(l)~\lambda(l)^2\\
\frac{d\lambda(l)}{dl}&=&-4~\omega(l)^2\lambda(l)\\
\frac{dv(l)}{dl}&=&-8~\omega(l)~\lambda(l)^2,
\end{eqnarray}
\end{mathletters}
where
\begin{mathletters}
\begin{eqnarray}
H(l)&=&\omega_{0}a^{\dagger}(l)a(l)+\lambda_{0}
~\left({a^{\dagger}(l)}^2+{a(l)}^2 \right)\nonumber\\
&=&\omega(l)a^{\dagger}a+\lambda(l)\left({a^{\dagger}}^2+a^2\right)
+v(l)
\end{eqnarray}
\begin{equation}
H_{d}(l)=\omega(l)a^{\dagger}a+v(l),
\end{equation}
and
\begin{equation}
a^{\dagger}(l)=U^{\dagger}(l)a^{\dagger}U(l),\quad a(l)=U^{\dagger}(l)aU(l)
\end{equation}
\end{mathletters}
with
$~\omega(0)=\omega_{0},~\lambda(0)=\lambda_{0},
~v(0)=v_{0}$.
One can easily find that 
the solution of these equations has satisfied
\begin{mathletters}
\begin{eqnarray}
&&\omega(l)^2-(2~\lambda(l))^2=\omega_{0}^2-(2~\lambda_{0})^2=const.\\
&&v(l)=v_{0}+\frac{1}{2}\left(\omega(l)-\omega_{0}\right).
\end{eqnarray}
\end{mathletters}
Notice that  there are two different type solutions whose behaviors are 
qualitatively different from each other (Fig.\ref{fig:unitary_flow}).
They can be distinguished by their initial points.

\noindent
(i) $|\lambda_{0}/\omega_{0}|<1/2$ case
\begin{mathletters}\label{eq:bounded_case}
\begin{eqnarray}
\omega(\infty)&\rightarrow&\omega_{0}\sqrt{1-4\left(\frac{\lambda_{0}}{\omega_{0}}\right)^2},\\
\lambda(\infty)&\rightarrow&0,\\
v(\infty)&\rightarrow&v_{0}
+\frac{1}{2}\left(\omega(\infty)-\omega_0\right),
\end{eqnarray}
\end{mathletters}
(ii) $|\lambda_{0}/\omega_{0}|>1/2$ (unbounded) case
\begin{mathletters}\label{eq:fixed_point_2}
\begin{eqnarray}
\omega(\infty)&\rightarrow&0,\\
\lambda(\infty)&\rightarrow&\lambda_{0}\sqrt{1-\frac{1}{4}\left(\frac{\omega_{0}}{\lambda_{0}}\right)^2},\\
v(\infty)&\rightarrow&v_{0}-\frac{1}{2}\omega_0 .
\end{eqnarray}
\end{mathletters}

Notice that the diagonal representation ($\lambda(\infty)=0$) is obtained 
in the case of (i) ($|\lambda_{0}/\omega_{0}|<1/2$).  
This diagonal representation (\ref{eq:bounded_case}) can be understood 
in terms of famous Bogoliubov transformation, i.e. 
\begin{mathletters}\label{eq:diagonalized_Hamiltonian}
\begin{equation}
H=\Omega b^{\dagger}b+\kappa+v_{0},\quad [b,b^\dagger]=1
\end{equation}
where
\begin{equation}
\Omega=\frac{\omega_{0}}{\sinh^2\theta+\cosh^2\theta},\quad
\kappa=-\frac{\omega_{0}\sinh^2\theta}{\sinh^2\theta+\cosh^2\theta}.
\end{equation}
\end{mathletters}
and 
\begin{mathletters}\label{eq:Bogoliubov_tr}
\begin{equation}
a=b \cosh \theta- b^{\dagger} \sinh \theta,\quad
a^{\dagger}=b^{\dagger} \cosh \theta- b \sinh \theta,
\end{equation}
with $\theta$ which satisfies
\begin{equation}\label{eq:Bogoliubov_theta}
\frac{\lambda_{0}}{\omega_{0}}=\frac{\sinh\theta\cosh\theta}
{\sinh^2\theta+\cosh^2\theta}.
\end{equation}
\end{mathletters}
One can easily check that relations (\ref{eq:bounded_case}) are completely
equivalent to (\ref{eq:diagonalized_Hamiltonian}), i.e.,
\begin{equation}
\omega(\infty)=\Omega,\quad v(\infty)=\kappa+v_0.
\end{equation}
The eigenvalues of (\ref{eq:hamiltonian}) can be easily obtained from  
(\ref{eq:bounded_case}),  that is
\begin{equation}
E_{n}=\Omega~n +\kappa+v_{0}\quad
(n ~\mbox{is integer}).
\end{equation}
As it should be, the original flow equation approach works in this case (i). 

In not only this case but also general cases where we can obtain a 
diagonal representation of a given Hamiltonian with this approach,  
the flow equation approach or the unitary transformation $U(\infty)$ 
can be interpretated as a transformation of the basis from one which 
diagonalizes $H_{d}(0)$ to another which diagonalize $H(0)$.  
We can say this original approach works when both of the basis 
are in a same function space because $U(\infty)$ is unitary.  
On the other hand, in the case of (ii)
it is clear that the eigenstates of the Hamiltonian is only scattering states
and they can not be in a same function space with 
eigenstates of $H_{d}(0)$ (boundary state). Therefore it is rather 
reasonable that the original approach does not work, as we saw in 
(\ref{eq:fixed_point_2}).

It would be obvious that some generalization to 
{\it larger class flow} should be required to obtain 
the diagonal representation of the Hamiltonian. 
The key should be to remove the unitarity 
from the flow equation approach and that it can be realized 
in the following procedure.

Generally, flows with the differential equation of the first-order 
like equation (\ref{eq:flow_equation}) does not have any crossing points 
with each other, {\it except for the attractor points}. Thus we would 
expect that some different flows 
whose attractor points are common will be related to each other. 
A diagonal representation of an 
unbounded Hamiltonian  is also obtained with some new flow 
which has same attractor points with the original (unitary) one, 
however this new flow would not represent an unitary 
transformation, unlike the original flow.  
General solutions of Eq.(\ref{eq:flow_equation}) 
which has a attractor point (\ref{eq:fixed_point_2}) 
can be written as 
\begin{mathletters}\label{eq:general_solution}
\begin{eqnarray}
\omega(l)&=&2\lambda(\infty)\left(\cos\varphi_l\sinh\xi_l
+i\sin\varphi_l\cosh\xi_l
\right),\\
\lambda(l)&=&
~\lambda(\infty)~\left(\cos\varphi_l\cosh\xi_l
+i\sin\varphi_l\sinh\xi_l\right),\\
v(l)&=&v_{0}+\frac{1}{2}\left(\omega(l)-\omega_{0}\right),
\end{eqnarray}
\end{mathletters}
where 
\begin{mathletters}\label{eq:C_flow_equation}
\begin{eqnarray}
\frac{d}{dl}\xi_l&=&-8\lambda(\infty) \cos 2\varphi_{l}\sinh 2\xi_{l},\\
\frac{d}{dl}\varphi_l&=&-8\lambda(\infty) \sin 2\varphi_{l}\cosh 2\xi_{l}
\end{eqnarray}
with
\begin{equation}
\xi_\infty\rightarrow 0,\quad \varphi_\infty\rightarrow 0.
\end{equation}
\end{mathletters} 
It should be noticed that the real (unitary) flow is one of the solutions 
represented with (\ref{eq:general_solution}), i.e., $\varphi_l=0$, and 
there are two other separable solutions 
i.e., $\varphi_l>0$ 
($\varphi_l\rightarrow +0$ as $l \rightarrow \infty$) and $\varphi_l<0$ 
($\varphi_l\rightarrow -0$ as $l \rightarrow \infty$).
which are originated from {\it unstable fixed-points} 
\begin{equation}\label{eq:unstable_fixed_points}
(\omega, \lambda)=(\pm 2i\lambda(\infty), 0),
~\left(\mbox{or} 
~(\xi,\varphi)=(0,\pm \frac{\pi}{2})\right).
\end{equation}
\begin{figure}
\epsfxsize=0.4\textwidth
\epsfbox{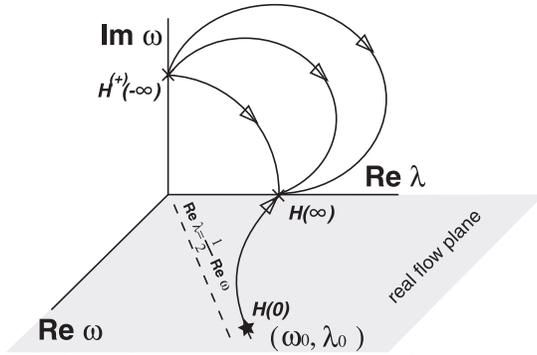}
\caption{Schematical illustration of the generalized flows $H^{(0)}(l)$ and 
$H^{(+)}(l)$. \label{fig:generalized_flow}}
\end{figure}
\noindent
There are three kinds of flows which have 
a common attractor point $\left(0,\lambda(\infty)\right)$ but 
different ``initial ($l\rightarrow-\infty$)" points, 
$\left(2i\lambda(\infty),0\right)$, 
$\left(-2i\lambda(\infty),0\right)$, and ``somewhere on the real flow plane".
Corresponding to them, we can introduce three 
Hamiltonian-flows, i.e., $H^{(\pm)}(l)$ and $H^{(0)}(l)$ as 
\begin{eqnarray}
H^{(j)}(l)&=&\omega^{(j)}(l)a^{\dagger}a
+\lambda^{(j)}(l)\left({a^{\dagger}}^2+a^2\right)+v^{(j)}(l)\\
(j&=&\{\pm,0\}),\nonumber
\end{eqnarray}
with
\begin{mathletters}\label{eq:boundary condition of the flow}
\begin{eqnarray}
\omega^{(\pm)}(-&&\infty)\rightarrow \pm 2i\lambda(\infty),
~\lambda^{(\pm)}(-\infty)\rightarrow 0,\nonumber\\
&&\left(v^{(\pm)}(-\infty)\rightarrow v(\infty)\pm i\lambda(\infty)\right)
\end{eqnarray}
and
\begin{eqnarray}
\omega^{(0)}(&&0)=\omega_{0},
~\lambda^{(0)}(0)=\lambda{_0},~\left(v^{(0)}(0)=v_{0}\right).
\end{eqnarray}
It is obvious that these Hamiltonian-flows are  attracted to 
the common fixed point (\ref{eq:fixed_point_2}) 
as $l\rightarrow \infty$, i.e.,
\begin{equation}\label{eq:attoractor_condition}
\omega^{(j)}(\infty)\rightarrow 0,
~\lambda^{(j)}(\infty)\rightarrow \lambda(\infty),
~v^{(j)}(\infty)\rightarrow v(\infty).
\end{equation}
\end{mathletters}
Precisely speaking, (\ref{eq:C_flow_equation}) and 
(\ref{eq:boundary condition of the flow}) are not sufficient 
to define $H^{(\pm)}(l)$ uniquely. They depend on  direction of an 
``infinitesimal shift'' from the fixed points.  
Giving this direction is giving some complex 
number $\varphi_{\infty}/\xi_{\infty}$. 
For example, $\varphi_{\infty}/\xi_{\infty}=0$ corresponds to $H^{(0)}(l)$.

The Hamiltonian flow $H^{(0)}(l)$ is 
the unitary flow described in (\ref{eq:fixed_point_2}), and that
not $H^{(0)}(l)$ but $H^{(\pm)}(l)$ give us the diagonal representation of 
the original Hamiltonian (\ref{eq:hamiltonian}) 
with complex eigenvalues (at $l\rightarrow -\infty$), i.e.
\begin{eqnarray}
H^{(\pm)}&&(-\infty)=\pm 2i\lambda(\infty)~a^{\dagger}a+v^{(\pm)}(-\infty)\nonumber\\
&&=\pm 2i\lambda_{0}\sqrt{1-\frac{1}{4}\left(\frac{\omega_{0}}{\lambda_{0}}
\right)^{2}}~\left(a^{\dagger}a+\frac{1}{2}\right)+v(\infty)\nonumber\\
&&\equiv \pm i\gamma ~\left(a^{\dagger}a+\frac{1}{2}\right)+v(\infty).
\label{eq:C_diagonal_representation}
\end{eqnarray}
We can immediately obtain the quantized complex eigenvalues
\begin{equation}\label{eq:C_eigenvalues}
E^{(\pm)}_{n}=\pm i{\gamma}\left(n +\frac{1}{2}\right)+v(\infty)
\quad (n ~\mbox{is integer}),
\end{equation}
and this is consistent with well-known 
result (for example, see \cite{ref:Shinbori-Kobayashi,ref:Shinbori}).
The symmetry in (\ref{eq:C_eigenvalues}) is related to symmetric 
phenomena in unstable system, that is,  
``decaying process for the positive time" 
and ``growing process for the negative time", as was pointed out 
in \cite{ref:Gamov,ref:Nakanishi,ref:Combes,ref:Bohm,ref:Shinbori-Kobayashi,ref:Shinbori}.  On should notice that the infinitesimal shift from 
the attractor point (\ref{eq:fixed_point_2}) decides
the arrow of time of the phenomena. 
This means that the structure of the generalized flow 
should be directly connected to the deep features of an irreversible
process which described with quantum mechanics. 

Let us mention to applications of our approach to other models. 

First, this approach seems possible to apply easily to other models, 
however we have to be careful of the following fact.
In general, when our  approach is applied to some systems, 
a truncation schemes to get a closed set 
of differential equations would be often required,  since 
the higher interactions will be generated  with the ``evolution'' $U(l)$.  
In such cases, we can adopt some truncation schemes which 
have already been proposed for the ordinary flow equation approach 
\cite{ref:Kehrein-Mielke,ref:Mielke} in which we can consider  
non-perturbative corrections.  

Second, in general cases it is almost impossible to solve the set of equations 
analytically because they becomes highly nonlinear even after the trancation.
Therefore, sometimes, 
we have to study generalized flow equations numerically.  
In the numerical studies, the following procedure to find an unstable 
fixed point (which gives a diagonal representation of unbounded Hamiltonian) 
would be useful.  
\begin{enumerate}
\item Find a stable fixed point 
(like (\ref{eq:fixed_point_2})) with the forward 
($l\rightarrow \infty$) evolution with 
the {\it unitary} flow from original (given) Hamiltonian.
This process should be always easy because the point is a attractor 
point. 
\item After finding the fixed point,  compute the {\it backward evolution} 
from the ``initial" point which is slightly (infinitesimally) shifted 
to some direction from the attractor point.  
\end{enumerate}
The flow 
will be ``attracted" to the unstable point as $l \rightarrow -\infty$, 
and we should easily find the unstable fixed points as  ``attractors". 
There must be such symmetric two unstable  points which 
depend on the direction of the  ``initial" infinitesimal shift. 
(This  corresponds to two Hamiltonian flows $H^{(\pm)}(l)$  
for (\ref{eq:hamiltonian}). )

Finally, let us mention to an application of our approach to 
a so-called open system problems.  As well known, there are two 
kinds of discussion about the time arrow in quantum mechanics \cite{ref:Bohm}. 
One is called the intrinsic irreversibility and 
the other is called the extrinsic irreversibility.
The decaying process of state with the Hamiltonian (\ref{eq:hamiltonian}) is 
thought as an example of the former, 
and the decay with an interaction with {\it environment} 
is thought as the latter or as a dynamics of the open system
\cite{ref:Feynman-Vernon,ref:Leggett}. 
Dealing with the open system is another standard approach to 
introduce the time arrow to quantum mechanics. 
It should be interesting problem 
to apply our approach to open systems.
Actually  spin-boson model 
which is a typical model to discuss the extrinsic decaying process
\cite{ref:Leggett}, was studied within an (improved but) unitary flow 
equation approach by Mielke {\it etal}~\cite{ref:Mielke}, 
and they succeeded in computing the energy shift 
of spin system which is due to the interaction with the 
bosonic environment. 
But  decay constants was not 
obtained directly and the dissipative property of the system 
seems hard to be understood  from their analysis.
We think our approach  can be used to reexamine this system and 
the dissipative properties of this system can be made clearest.
Moreover it will be helpful to understand 
the relation between two different treatment on the irreversibility in 
quantum mechanics. 


Kentaro Imafuku is grateful to Centro Vito 
Volterra and Luigi Accardi for kind hospitality.
He is supported by a overseas research fellowship of Japan Science and 
Technology Corporation. 


\end{multicols}
\end{document}